\documentstyle[seceq,epsf,wrapft,twoside]{ptptex}
\notypesetlogo  
\title{Relation between Scattering and Production Amplitudes\\
--- Case of $\sigma$ and $\kappa$ in $\pi\pi$ and $K\pi$ Systems --- 
}
\author{%
Muneyuki {\sc Ishida} and Shin Ishida$^*$
}
\inst{%
Department of Physics, Meisei University, 
Hino 191-8506, Japan\\
$^*$  Research Institute of Quantum Science, 
College of Science and Technology\\
Nihon University, Tokyo 101-0062, Japan
}
\abst{%
In our talk at Nihon university symposium, 
we have argued that
the $\pi\pi / K\pi$ production amplitudes ${\cal F}$ generally have different structures
from those of the $\pi\pi / K\pi$ scattering amplitudes ${\cal T}$
from a viewpoint of generalized $S$ matrix on the right bases, reflecting the quark-physical picture of strong
interactions, and that 
${\cal F}$ should be analyzed independently from ${\cal T}$.
Here we discuss the miscellaneous topics related with this argument:
The conventional analyses by ${\cal K}$ matrix method, 
which necessarily lead to the common phases and structures through both ${\cal T}$
and ${\cal F}$,
are based on improper application of elastic unitarity to the production processes. 
Recent criticisms on the method of analyses of $J/\psi \rightarrow \omega\pi\pi$ and 
$D^+\rightarrow K^-\pi^+\pi^+$, which led to the existence of $\sigma$ and $\kappa$, respectively, 
are also based on improper application of elastic unitarity.
}

\begin{document}
\maketitle

\setcounter{tocdepth}{4}

\section{Introduction}

In our talk\cite{II} at Nihon university referred to as II, 
we have investigated the relation between the $\pi\pi / K\pi$ production amplitude ${\cal F}$ 
and the scattering amplitude ${\cal T}$ from a viewpoint of unitarity of generalized $S$ matrix,

For many years it had been believed that both ${\cal T}$ and ${\cal F}$
in the low mass region with $\sqrt s < 1$GeV have the same structures 
(universality of the scattering amplitude\cite{pen}), and that analyses of any production process 
should be done together with the scattering process. 
Especially, regardless of the fact that ${\cal F}$ often has the peak structure clearly 
due to $\sigma /\kappa$ meson, 
it is believed\cite{bug} to show the same phase motion $\stackrel{<}{\scriptstyle \sim}90^\circ$ 
as ${\cal T}$ because of the elastic unitarity (or final state interaction theorem). 
Because of this belief the observed peak structures in the various production processes had been 
regarded as mere backgrounds. 

However, in II we have argued that the elastic unitarity does not work in the relevant  
$J/\psi$ and $D$ decays. The above belief is not correct and  
the production processes should be independently\cite{Sawazaki,ShinMune,SUNY} treated from the scattering process.
The essential reason of this is that as the basic fields of expanding $S$ matrix, 
the ``bare" fields of $\sigma$ and $\kappa$, as well as the $\pi$ and $K$ fields, 
should be taken into account, 
since all these states are equally color-singlet bound states of quarks. 

In this talk, we first review the above argument in \S 2.
The notion of generalized $S$ matrix and its unitarity are explained in some details. 
The miscellaneous topics are treated in \S 3, where 
the criticisms, concerning the unitarity and phase motion of our ${\cal F}$, presented recently, 
are discussed:
In the conventional analyses by ${\cal K}$ matrix method, 
the phases of ${\cal F}$ are constrained to be the same as the scattering phase shift,
and the ${\cal F}$ are commonly fitted to ${\cal T}$.
However, as will be explained in \S 3, this constraint is based on 
improper application of elastic unitarity to the production processes. 
The recent criticisms concerning the phase motion of ${\cal F}$ for 
$J/\psi \rightarrow \omega\pi\pi$ and $D^+\rightarrow K^-\pi^+\pi^+$
will also be clarified not to be valid.

\section{Unitarity of Generalized $S$ Matrix}

As was discussed in II,
the strong interaction is a residual 
interaction of QCD among all color-singlet bound states of quarks.
These states are denoted as $\phi_i$. 
The set of $\phi_i$ fields is considered to include 
the $\sigma$ and $\kappa$, as well as the $\pi$ and $K$ states, 
since all these states are equally to be the bound staes of quarks.
The strong interaction Hamiltonian ${\cal H}_{\rm str}$ is described by the 
$\phi_i$ fields: ${\cal H}_{\rm str} = {\cal H}_{\rm str}  (\phi_i) $.
The time-evolution by ${\cal H}_{\rm str} (\phi_i)$ describes the generalized $S$-matrix,
and the bases of generalized $S$-matrix\cite{ShinMune,SUNY,II} are considered 
to be the configuration space of these multi-$\phi_i$ states. 

Here, it is to be noted that, if ${\cal H}_{\rm str}$ is hermitian, the unitarity of 
$S$ matrix is guaranteed.
 
By taking the relevant $J/\psi\rightarrow\omega\pi\pi$ decay as an example,
 $| \omega\sigma \rangle$, $| \omega f_2 \rangle$, $| b_1 \pi \rangle$, $|J/\psi\rangle$,
 $| \omega \pi\pi \rangle$, $\cdots$ are necessary bases to describe this process(,
since the $\sigma$, $f_2$ $\cdots$ successively decay to $\pi\pi$ and the $b_1$ decays to $\omega\pi$, 
and the final states become $\omega\pi\pi$). 
The generalized $S$-matrix relevant to this process is given by
\begin{footnotesize}
\begin{eqnarray}
 && \begin{array}{ccccc} \ \ \ \ \ \ |J/\psi\rangle_{in} &  \ \ \ \ \ \ \ \ |\omega\sigma\rangle_{in} &
             \ \ \ \  \ \ \ \  |\omega f_2\rangle_{in} &   \ \ \ \  \ \ \ \  |b_1\pi \rangle_{in} & 
           \ \ \ \  \ \ \ \   |\omega \pi\pi \rangle_{in} \ \ \cdots\\  \end{array} \nonumber\\
\begin{array}{c}     {}_{out}\langle J/\psi |  \\   
      {}_{out}\langle \omega\sigma |  \\ 
      {}_{out}\langle \omega f_2 | \\  
      {}_{out}\langle b_1 \pi | \\   \\    
     \end{array}     
 &&
\left( \begin{array}{ccccc} \sqrt{1-\eta^2_{\psi}}e^{2i\theta_{\psi}} &  
                   r_{\psi \sigma} e^{i \theta_{\psi \sigma} } &  r_{\psi f_2} e^{i \theta_{\psi f_2}} &  
                    r_{\psi b_1} e^{i \theta_{\psi b_1}}  &  \cdots   \\
 r_{\psi \sigma} e^{i \theta_{\psi \sigma} }  &  \sqrt{1-\eta^2_{\sigma}}e^{2i\theta_{\sigma}} &   
                   r_{\sigma f_2} e^{i \theta_{\sigma f_2}} &   r_{\sigma b_1} e^{i \theta_{\sigma b_1}}  &  \cdots   \\
 r_{\psi f_2} e^{i \theta_{\psi f_2}} &    r_{\sigma f_2} e^{i \theta_{\sigma f_2}} &  \sqrt{1-\eta^2_{f_2}}e^{2i\theta_{f_2}} &   
           r_{f_2 b_1} e^{i \theta_{f_2 b_1}}  &  \cdots   \\
 r_{\psi b_1} e^{i \theta_{\psi b_1}} &  r_{\sigma b_1} e^{i \theta_{\sigma b_1}} &   r_{f_2 b_1} e^{i \theta_{f_2 b_1}} &   
                  \sqrt{1-\eta^2_{b_1}}e^{2i\theta_{b_1}}  &  \cdots   \\
\vdots & \vdots & \vdots & \vdots  & \ddots \\  
\end{array} \right)  \ 
\label{c2.2}
\end{eqnarray}
\end{footnotesize}
where $|\ \ \rangle_{in}$ and  ${}_{out}\langle \ \ |$ denotes $in$ and $out$-state of 
scattering theory, and 
$r_{\psi\sigma}e^{i\theta_{\psi\sigma}} = {}_{out}\langle \omega\sigma | J/\psi\rangle_{in}$
for example. The $n\times n$ $S$ matrix, 
satifying unitarity ($SS^\dagger =1$) and time-reversal invariance ($S_{ij}=S_{ji}$), 
has $\frac{n(n+1)}{2}(=n+\frac{n(n-1)}{2})$ independent parameters,
which correspond to the phases of diagonal elements 
($\delta_{\psi}$, $\delta_\sigma$, $\cdots$; $n$ freedoms) and the couplings of non-diagonal 
elements ($r_{\psi\sigma}$, $r_{\sigma f_2}$, $\cdots$; $\frac{n(n-1)}{2}$ freedoms). 
The other quantities in  Eq.~(\ref{c2.2}) is represented by these $\frac{n(n+1)}{2}$ parameters: 
For example 
 $\theta_{\psi\sigma}=\theta_{\psi}+\theta_{\sigma}+\phi_{\psi\sigma}$(, where 
the $\phi_{\psi\sigma} =\phi_{\psi\sigma} (r_j)$ is determined only from the 
non-diagonal coupling parameters $r_j(j=\psi\sigma , \psi f_2 , \sigma f_2 ,\cdots )$
due to unitarity constraint). The VMW method is based on this $S$-matrix parametrization method 
satisfying unitarity(generalized unitarity). 


The effective $\omega\pi\pi$ amplitude ${\cal F}_{\omega \pi \pi}$ is given by 
a coherent sum of the elements in the first column of $S$, Eq.~(\ref{c2.2}),
where the decays of $\sigma$, $f_2$, $\cdots$ to $\pi\pi$ and that of $b_1$ to $\omega\pi$ are
described by Breit-Wigner formulas, $\Delta (s)=M \Gamma /(M^2 - s - i M \Gamma )$. 
\begin{eqnarray}
  {\cal F}_{ \omega  \pi  \pi} \ \ \  &=& 
   {\cal F}_{\omega\sigma}+ {\cal F}_{\omega f_2}  
   +{\cal F}_{ b_1\pi} +\cdots + {\cal F}_{\omega (\pi\pi)_{Non.Res.}} ,
\label{c4}\\
 {\cal F}_{\omega\sigma}\ \ \ 
     &=& {}_{out}\langle \omega\sigma | J/\psi \rangle_{in}\Delta_\sigma (s) 
     = r_{\psi\sigma} e^{i\theta_{\psi\sigma}}  
       \frac{m_\sigma \Gamma_\sigma}{m_\sigma^2-s-im_\sigma\Gamma_\sigma (s)} \nonumber\\
{\cal F}_{\omega f_2}\ \ \  
     &=& {}_{out}\langle \omega f_2 |\  J/\psi\  \rangle_{in} \Delta_{f_2}(s)
     = r_{\psi f_2} e^{i\theta_{\psi f_2}} 
       \frac{m_{f_2} \Gamma_{f_2} N_{\pi\pi}(s,{\rm cos}\theta)}
              {m_{f_2}^2-s-im_{f_2}\Gamma_{f_2} (s)} \nonumber\\
{\cal F}_{\omega b_1}\ \ \  
     &=& {}_{out}\langle \omega b_1 |J/\psi \rangle_{in} \Delta_{b_1}(s) 
     =r_{\psi b_1} e^{i\theta_{\psi b_1}} 
        \frac{m_{b_1} \Gamma_{b_1}}
              {m_{b_1}^2-s-im_{b_1}\Gamma_{b_1} (s)}  \nonumber\\
     \cdots &&  \nonumber\\
  {\cal F}_{\omega (\pi\pi)}^{Non.Res.} 
     &=& {}_{out}\langle \omega (2\pi)_{N.R.} | J/\psi \rangle_{in} \ \ \ \ \ 
     =\ \ \ r_{2\pi}^{N.R.} e^{i\theta_{2\pi}^{N.R.}}\ \ .
\label{c5}
\end{eqnarray}
Each term of ${\cal F}$ has mutually-independent respective couplings, 
$r_{\psi\sigma}$, $r_{\psi f_2}$, $r_{\psi b_1}$, $\cdots$ 
and $r_{\psi2\pi}^{N.R.}$. 
The phases $\theta_{\psi\sigma}$, $\theta_{\psi f_2}$, $\cdots$ are related with 
the phases of diagonal elements of $S$, $\theta_{\psi}$, $\theta_{\sigma}$, $\cdots$, which are unknown.
Thus, we may treat phenomenologically the formers as independent parameters from the latters.    
The amplitude given by Eqs.~(\ref{c4}) and (\ref{c5}) is consistent with the $S$-matrix 
unitarity(generalized unitarity). This parametrization method is called as VMW method.

We should note that the imaginary parts, $-iM\Gamma$, of the denominators of $\Delta (s)$ for $\sigma$, $f_2$, $\cdots$
does not come from the $\pi\pi$ final state interaction or the virtual $\pi\pi$-loop effect, but does
from their real decay processes to $\pi\pi$.\footnote{
In non-relativistic quantum mechanics, the decay of the unstable particle can be described 
by the WF with imaginary part in the energy, 
$e^{-iE_0t}\rightarrow e^{-i(E_0-i\Gamma /2)t}$. Correspondingly, the propagator is replaced as 
$\frac{1}{E_0-E}\rightarrow \frac{1}{E_0 - E - i \Gamma /2}$ .
Similarly, in the Feynman propagator, 
$\Delta_F(x)=\int \frac{d^4k}{(2\pi )^4}\frac{e^{ikx}}{m^2+k^2-i\epsilon}
=\int \frac{d^3{\bf k}}{i(2\pi )^32\omega}e^{-i\omega |x_0|}e^{i{\bf k}\cdot{\bf x}} $, 
when we replace 
 $e^{-i\omega |x_0|}$ by  $e^{-i (\omega -i\Gamma /2)|x_0|}$, we obtain the propagator
$\frac{1-i\Gamma /(2\omega )}{m^2+k^2-\Gamma^2/4-i\omega\Gamma}$, which may be approximated as
$\frac{1}{m^2-s-i m \Gamma (s)}$ in $\Delta (s)$.
}  

Corresponding to this argument, we should further note the following point:
In the conventional method of analyses of the relevant 3-body decays, 
$J/\psi\rightarrow \omega\pi\pi$\cite{DM2,had97Jpsi,WuNing} and $D^-\rightarrow\pi^-\pi^+\pi^+$\cite{E791}, 
the $\pi\pi$ system in low $m_{\pi\pi}$ region is believed to decouple from the remaining particle
$\omega$ and $\pi^+$ in the final channel. Accordingly, 
because of the Watson final state interaction theorem (or $\pi\pi$ elastic unitarity),
the amplitudes ($\ref{c4}$) are to take the same phase as the $\pi\pi$ scattering 
phase shift $\delta$. 
However, on the contrary to this conventional belief, 
experimentally the relevant $\pi\pi$ system seems to show the behavior not
to be isolated in the final channel.
Actually, the large strong phases are suggested experimentally in $J/\psi$ decays and $D$ decays:
In $J/\psi\rightarrow 1^-0^-$ decays 
(that is, $J/\psi\rightarrow \omega\pi^0,\rho\pi,K^*\bar K,\cdots$,)\cite{Mahiko} 
it is necessary to introduce a large relative strong phase\cite{Mahiko} 
$\delta_\gamma =arg\frac{a_\gamma}{a}=80.3^\circ$ 
between the effective coupling constants
of three gluon decay $a$ and of one photon decay $a_\gamma$.
A similar result is also obtained in  $J/\psi\rightarrow 0^-0^-$ decays.
A large relative phase between $I=3/2$ and $I=1/2$ amplitudes 
of $D\rightarrow K\pi$ decays is observed: 
$\delta_{3/2}(m_D)-\delta_{1/2}(m_D)=(96\pm 13)^\circ$,\cite{Dphase}
(while in $B\rightarrow D\pi ,D\rho ,D^*\pi$ decays rather small relative phases are obtained).
According to this work, $M_\psi$($M_D$) is not sufficiently large 
for making $\pi\pi$ decouple from $\omega$(the other $\pi^+$), and the $\pi\pi$ elastic 
unitarity constraint actually does not work in the amplitude Eq.~(\ref{c4})
of $J/\psi\rightarrow \omega\pi\pi$ and $D \rightarrow \pi^-\pi^+\pi^+$.
Similar consideration is also applicable to the 
$J/\psi\rightarrow K^* K\pi$\cite{WNkappa,Komada,WNII} and $D \rightarrow K^-\pi^+\pi^+$\cite{E791kappa}.
The final $K\pi$ does not decouple from $K^*$ and  the other $\pi^+$. The elastic unitarity does not work
and the corrsponding amplitude ${\cal F}$ takes different phase from the $K\pi$ scattering phase shift.  

In the amplitude (\ref{c4}) and (\ref{c5}) we use the $\sigma$ Breit-Wigner formula 
having the phase motion by 180 degrees, which are different from the motion of 
$\delta$ about 90 degrees below 0.9GeV.  
In the relevant $J/\psi$ decays and $D$ decays, the low $m_{\pi^+\pi^-}$ regions  
are dominated by $\sigma$, and the Breit-Wigner phase motion of $\sigma$ are expected to be observed.
In the recent experimental study\cite{brazil,Gobel} of  $D^+\rightarrow \pi^-\pi^+\pi^+$ decays,
 180$^\circ$ phase motion of $\sigma$-meson is actually observed.
This fact clearly shows the validity of our consideration, that is,
the $\sigma$ is not the scattering state formed from final $\pi\pi$ interaction
but corresponds to the members of the general $S$-matrix bases.

\section{Miscellaneous Topics: Reply to the Recent Criticisms}

As was explained in the previous section, 
our methods of analyses applying Eq.~(\ref{c4}) are based on the unitarity
of generalized $S$ matrix.
Recently the following criticisms to them, concerning the 
phase motion of ${\cal F}$, have been presented. Here we reply to them. 

({\it cos $\theta$ distribution in $J/\psi\rightarrow\omega\pi\pi$})\ \ \ \ 
Minkowski and Ochs raise a criticism 
concerning the existence of light-mass $\sigma$-pole\cite{MO1} 
in $J/\psi\rightarrow\omega\pi\pi$:\ \ \ 
The cos$\theta$ distribution is obtained 
in $m_{\pi\pi}$=250$\sim$750MeV~by~DM2.\cite{DM2}
They apply partial wave expansion(PWA) including $S$ and $D$ waves 
to obtain the cross section as
$
\frac{d\sigma}{d\Omega} \sim |S|^2 + 10 ( 3 {\rm cos}^2\theta -1) Re(SD^*) +{\cal O}(|D|^2),
$
($S(D)$ is the $\pi\pi$ S(D)-wave component), where 
the cos$^2\theta$ term is proportional to $Re(SD^*)$.
Then, if the $D(S)$ is dominated by $f_2(1270)\ (\ \sigma\ )$ contribution,
the angular distribution would vary with a sign change of 
the cos$^2\theta$ term (from $+$ to $-$).
Actually, contradictorily to this anticipation, 
the data do not show any sign change below 750 MeV,
and they conclude that there is no indication for
a Breit-Wigner resonance at 500 MeV.\cite{MO1}

However, (according to our preliminary analyses,\cite{sigmaG}) in this energy region 
there is almost no contribution from $f_2(1275)$, and 
actually the $l\geq 2$ partial waves mainly come from $b_1(1235)$ contribution,
$J/\psi\rightarrow \pi b_1$ and $b_1\rightarrow \omega\pi$.
In $m_{\pi\pi}\stackrel{>}{\scriptscriptstyle \sim} 500$MeV the direct $b_1$ peak 
is seen in cos $\theta$ distribution, and it includes
the large higher wave components.
This fact means the above  PWA does not work well in this energy region. 
Furthermore, each of the partial waves is expected to show 
a large phase movement (from $b_1$ pole)
in the relevant energy region $m_{\pi\pi}\sim 500$MeV, while in the above criticism
the almost constant phase of $D$ wave is assumed.  
Thus, the basic assumption is not applicable,
and their criticism is not correct. 

Recently the 180 degrees phase motion of $\sigma$ is observed 
in $D^+\rightarrow\pi^-\pi^+\pi^+$ decays by using the interference of 
$\sigma$ and $f_2(1275)$.\cite{brazil,Gobel}
Similarly, in $J/\psi\rightarrow\omega\pi\pi$, it may be possible to observe 
the $\sigma$ phase motion by using the interference between $b_1(1235)$ Breit-Wigner amplitude 
and $\pi\pi$ $S$-wave component in Dalitz plot data.      
The direct $b_1$ peak appears in $m_{\pi\pi}\stackrel{>}{\scriptscriptstyle \sim} 500$MeV region
of Dalitz plot, and in this energy region the $S$-wave phase motion is expected to be determined 
with good accuracy.\footnote{In case $(m_\sigma ,\Gamma_\sigma )=(500,350)$MeV the $\sigma$
Breit-Wigner amplitude gives the phase difference $\Delta\delta =83^\circ (67^\circ)$
between $m_{\pi\pi}$=450$\sim$850MeV(500$\sim$850MeV) which is somewhat larger than 
the corresponding $\pi\pi$ scattering phase difference $\Delta\delta \simeq 63^\circ (55^\circ)$.}

({\it $D^+\rightarrow K^-\pi^+\mu^+\nu$ and $D^+\rightarrow K^-\pi^+\pi^+$ })\ \ \ \ 
The $K^-\pi^+$ spectra of $D^+\rightarrow K^-\pi^+\mu^+\nu$ by FOCUS\cite{FOCUS} is 
dominated by $P$ wave from $\bar K^{*0}$ interfering with a small
$S$ wave. Through the angular analysis this $S$ wave component
has almost constant phase $\delta =\frac{\pi}{4}$ 
in mass region of $\kappa$ meson,
$m_{K\pi}=0.8\sim 1.0$GeV. 
This $\delta$ is suggested, by Minkowski and Ochs\cite{MO1},
to be the same as the $K\pi$ scattering phase shift\cite{LASS} by LASS in this mass region.
Based on this result, they critisize the analysis of $D^+\rightarrow K^-\pi^+\pi^+$ 
by E791\cite{E791kappa}, where, as is explained in \S2, the $\kappa$ Breit Wigner amplitude 
is applied and it has large phase motion in this mass region. 
They stated 
``Such a result (of E791) appears to contradict the above FOCUS result.''

However, this criticism is again premature because the effect of strong phases
allowed in generalized unitarity condition is overlooked.
In $D^+\rightarrow K^-\pi^+\mu^+\nu$ the final $K^-\pi^+$ is isolated 
in strong interaction level.
Thus, as we explained in the first part of this section, 
the amplitude has the same phase as
$K\pi$ scattering amplitude due to Watson theorem.
However,  $D^+\rightarrow K^-\pi^+\pi^+$ is a decay into three strongly interacting hadrons
of heavy meson
and $K^-\pi^+$ is not isolated in the final channel, and the rescattering effects
from various elements of generalized $S$ matrix, such as, $_{out}\langle\bar\kappa^{0}\pi^+|J/\psi\rangle_{in}$,
$_{out}\langle\bar K^{*0}\pi^+|J/\psi\rangle_{in}$, $_{out}\langle\bar K_2^{*0}\pi^+|J/\psi\rangle_{in}$,
$\cdots$, must be taken into account. 
In two-body decays of $D$, $J/\psi$ ($B$) mesons, the large (small) strong phase is 
suggested.\cite{Mahiko} Thus, the phases of the above matrix elements
are considered to be not small.
Thus, 
total amplitude of $D^+\rightarrow K^-\pi^+\pi^+$ 
has generally different phase from that of 
$D^+\rightarrow K^-\pi^+\mu^+\nu$.

We should add a comment on $D^+\rightarrow K^-\pi^+\mu^+\nu$ 
in relation to $K_{l4}$ decay $K^+\rightarrow \pi^+\pi^- e^+\nu$.
The latter process is analyzed by Shabalin\cite{Shabalin} by using SU(3) linear $\sigma$ model;
where the effect of direct $\sigma$ production, 
$K^+\rightarrow \sigma e^+\nu $ 
(successively $\sigma\rightarrow \pi^+\pi^-$), explains 
the large width obtained experimentally 
(which is twice as large as the prediction by soft pion limit), and
at the same time this amplitude has the same phase 
as the $\pi\pi$ scattering phase shift. 
{\it The $\sigma$ Breit-Wigner phase motion is not observed 
due to Watson theorem, while
its large decay width suggests the $\sigma$ production} 
in this process. As can be seen in this example,
the $\kappa$ phase motion is not observed in  
$D^+\rightarrow K^-\pi^+\mu^+\nu$, but this fact
does not mean no $\kappa$-existence. 
The analysis of $D^+\rightarrow K^-\pi^+\pi^+$ by E791 does not contradict 
with FOCUS result, and strongly suggests the $\kappa$(800) existence.  

(${\cal K}$-$matrix$ $method$ $for$ $\pi\pi (K\pi)$-$production$)\ \ \ \  
The ${\cal K}$-matrix parametrization method is frequently\cite{pen,CB,Zou,Anisovich,Malvezzi} used for the analyses of 
$\pi\pi$ (or $K\pi$) scattering process.
In the most simple one($\pi\pi$)-channel case which may be applicable below $K\bar K$ threshold,
the scattering amplitude ${\cal T}$ is represented by
\begin{eqnarray}
{\cal T} &=& {\cal K}/(1- i \rho {\cal K}),\ \ 
  {\cal K}=\frac{g^2}{m_0^2-s} + \lambda_{NR} = (s-s_0)\frac{\hat g^2}{m_0^2-s}\ , 
\label{d1}
\end{eqnarray}
where the ${\cal K}$-matrix consists of a resonant part $g^2/(m_0^2-s)$ and 
of non-resonant part $\lambda_{NR}$.
They cancel with each other and lead to the Adler zero factor $(s-s_0)$ in the final form. 
This factor reproduces the threshold suppression of spectra of ${\cal T}$ consistently with chiral symmetry, 
and at the same time the slowly rising phase motion below 0.9GeV.
The ${\cal K}$-matrix method is also applied\cite{Anisovich,Malvezzi} to 
the various $\pi\pi$ (or $K\pi$) production processes, where the production 
amplitude ${\cal F}$ is represented by 
\begin{eqnarray}
{\cal F} &=& {\cal P}/(1-i\rho {\cal K}),\ \ 
   {\cal P}=\frac{e^{i\theta_0} \xi_0 g}{m_0^2-s} + e^{i\theta_{NR}}\xi_{NR}\ .\ \   
\label{d2}
\end{eqnarray}
Here, in the numerator ${\cal P}$, the production couplings $\xi_0$, $\xi_{NR}$  
and phases $\theta_0$, $\theta_{NR}$
independent from the scattering process are taken into account, while the denominator $1/(1-i\rho {\cal K})$ is 
the same as that of ${\cal T}$ in Eq.~(\ref{d1}). 
Correspondingly all the denominators give to total ${\cal F}$ the 
common phase motion (aside from the effects due to phases introduced in the numerator)
as ${\cal T}$, and it is insisted\cite{bug} that  
the common fit to the scattering and production processes are necessary. 
The imaginary part of the denominator in ${\cal F}$, Eq.~(\ref{d2}), has 
the $(s-s_0)$ factor, which is called Adler 0 in Ref.~\citen{bug}.

However, {\it such a requirement on ${\cal F}$ has no relation with
the Adler self-consistency condition which predicts the zero in total ${\cal F}$.
Moreover, that is not generally valid since in its approach only the $\pi\pi$
dynamics, that is, the final state interaction between two stable $\pi$ mesons
is considered, and the other various final state interactions
are not taken into account.}
Actually it is now confirmed experimentally that ${\cal F}$ of the amplitude of $D$ decays have
different phases from ${\cal T}$.\cite{Gobel} 
In the case of $J/\psi\rightarrow\omega\pi\pi$, the emitted pion energy is of order $M_\psi$,
and the new dynamics in $J/\psi$ energy region, which is beyond the scope
of chiral dynamics, must also be considered.
This is overlooked\cite{bug} in Eq.~(\ref{d2}).
The ${\cal K}$-matirix method, Eq.~(\ref{d2}), is not applicable to the analyses of $D$-decays and $J/\psi$ decays,
where the amplitudes ${\cal F}$ should\cite{Tuan} be independently parametrized from ${\cal T}$ by using VMW method.

\section{Concluding Remarks}

We have investigated the relation between the $\pi\pi /K\pi$ production amplitudes ${\cal F}$
and the scattering amplitudes ${\cal T}$ from the viewpoint of unitarity of generalized $S$ matrix.
From the quark physical picture we should treat $\sigma$ and $\kappa$ states as well as $\pi$
and $K$ states, both of which are bound states of quarks, on the same footing, and these states are
considered to form the bases of generalized $S$ matrix.
The ${\cal F}$ correspond to the different elements of generalized $S$ matrix from ${\cal T}$,
and should be analyzed independently from ${\cal T}$ in principle.

On the contrary, for these two decades, the so-called ``universality" of $\pi\pi / K\pi$ scattering,
which is based on the elastic unitarity, had been regarded seriously,
and the ${\cal F}$ were analyzed commonly with ${\cal T}$,
using the ${\cal K}$ matrix method.
However, as is explained in this talk, the validity of elastic unitarity is very limited.
Accordingly the most analyses along the line of thought on the above ``universality"
are considered to be revised.

Finally we should like to point out that, in the
relevant heavy meson decays, 
the $\pi\pi /K\pi$ systems 
are in the region of more high energy, where 
the various new physics\cite{PLB2}, other than the low energy dynamics,
should be generally taken into account.

\acknowledgements

The authors express their thanks to Prof. S. F. Tuan for useful information.
They are also grateful to Prof. K. Takamatsu, 
Prof. T. Tsuru and Prof. K. Ukai for discussions.

\end{document}